\documentclass[aps,prl,twocolumn,superscriptaddress]{revtex4-2}
\usepackage[english]{babel}
\usepackage{amsmath,amsthm,lipsum}
\usepackage{amsfonts}
\usepackage[normalem]{ulem}
\usepackage[pdfborder={0 0 0}, colorlinks=true, urlcolor=blue,linkcolor=blue, citecolor=blue]{hyperref}
\usepackage{color}
\usepackage{graphicx}
\usepackage{epstopdf}
\usepackage{float}
\usepackage{subfigure}
\usepackage{amsfonts}
\usepackage{mathrsfs}

\newcommand{\Grep}{\mathscr{G}}  
\usepackage{url}

\makeatletter
\newsavebox{\@brx}
\newcommand{\llangle}[1][]{\savebox{\@brx}{\(\m@th{#1\langle}\)}%
  \mathopen{\copy\@brx\kern-0.5\wd\@brx\usebox{\@brx}}}
\newcommand{\rrangle}[1][]{\savebox{\@brx}{\(\m@th{#1\rangle}\)}%
  \mathclose{\copy\@brx\kern-0.5\wd\@brx\usebox{\@brx}}}
\makeatother

\newcommand{\ba}{\begin{eqnarray}}
\newcommand{\ea}{\end{eqnarray}}

\theoremstyle{definition}

\theoremstyle{remark}

\begin{document}
\title{Emergence of Triplet Superconductivity from Cavity Vacuum Fluctuations} 
\author{Xin-Xin Yang}
\affiliation{Shanghai Qizhi Institute and Shanghai Artificial Intelligence Laboratory, Xuhui District, Shanghai 200232, China}
\affiliation{State Key Laboratory of Surface Physics, Institute of Nanoelectronics and Quantum Computing, Department of Physics, Fudan University, Shanghai 200438, China}

\author{Shuai Zhang}
\affiliation{State Key Laboratory of Surface Physics, Institute of Nanoelectronics and Quantum Computing, Department of Physics, Fudan University, Shanghai 200438, China}

\author{Kun Ding}
\thanks{kunding@fudan.edu.cn}
\affiliation{State Key Laboratory of Surface Physics, Institute of Nanoelectronics and Quantum Computing, Department of Physics, Fudan University, Shanghai 200438, China}

\author{Xiaopeng Li}
\thanks{xiaopeng\underline{ }li@fudan.edu.cn}
\affiliation{State Key Laboratory of Surface Physics, Institute of Nanoelectronics and Quantum Computing, Department of Physics, Fudan University, Shanghai 200438, China}
\affiliation{Shanghai Qizhi Institute and Shanghai Artificial Intelligence Laboratory, Xuhui District, Shanghai 200232, China}
\affiliation{Shanghai Research Center for Quantum Sciences, Shanghai 201315, China} 
\affiliation{Hefei National Laboratory, Hefei 230088, China}
\date{\today }

\begin{abstract}
Engineering quantum materials with cavity fields has emerged as a powerful route to manipulate phases of quantum matter in solids. 
Here we demonstrate that cavity vacuum fluctuations alone can drive the emergence of triplet superconductivity in an otherwise singlet superconductor.  The vacuum field renormalizes the electronic band structure in a polarization-dependent manner, reshaping the Fermi surface and altering the competition among symmetry-allowed pairing channels. As a result, multiple superconducting phases arise from the cavity vacuum fluctuations. Above a critical light–matter coupling, the leading instability switches from singlet to triplet pairing, yielding a superconducting state absent in the bare material. This vacuum-induced symmetry transition produces distinct modifications of the gap structure and low-energy quasiparticle spectrum. Our results establish cavity vacuum engineering as a mechanism for generating unconventional superconducting phases and stabilizing triplet states of potential relevance for topological superconductivity.
\end{abstract}

\maketitle

{\it Introduction.--}
Cavity materials engineering extends cavity quantum electrodynamics to solids and offers a route to tune material properties beyond conventional control parameters~\cite{Lu2025,Garcia2021,Bloch2022,Schlawin2022,Bretscher2026,Baydin2025}. It is now being explored across a broad range of condensed-matter phenomena, including Landau-level renormalization and quantum Hall physics~\cite{Li2018,Paravicini2019,Enkner2024,Enkner2025,Appugliese2022,Xue2025,Wang2019}, flat-band physics~\cite{Jiang2024,Liu2025}, topological phases~\cite{Liu2025,Lin2023,Nguyen2023,Dag2024}, and magnetic phase transitions~\cite{Latini2021,Ashida2020,Curtis2023}, among others~\cite{Jarc2023,Masuki2024,Chiocchetta2021,Tay2025,Riolo2025}. These advances establish cavity control as a general strategy for steering emergent phases in solids and naturally bring superconductivity into focus.

\begin{figure}[htbp]
    \centering
    \includegraphics[width=0.5\textwidth]{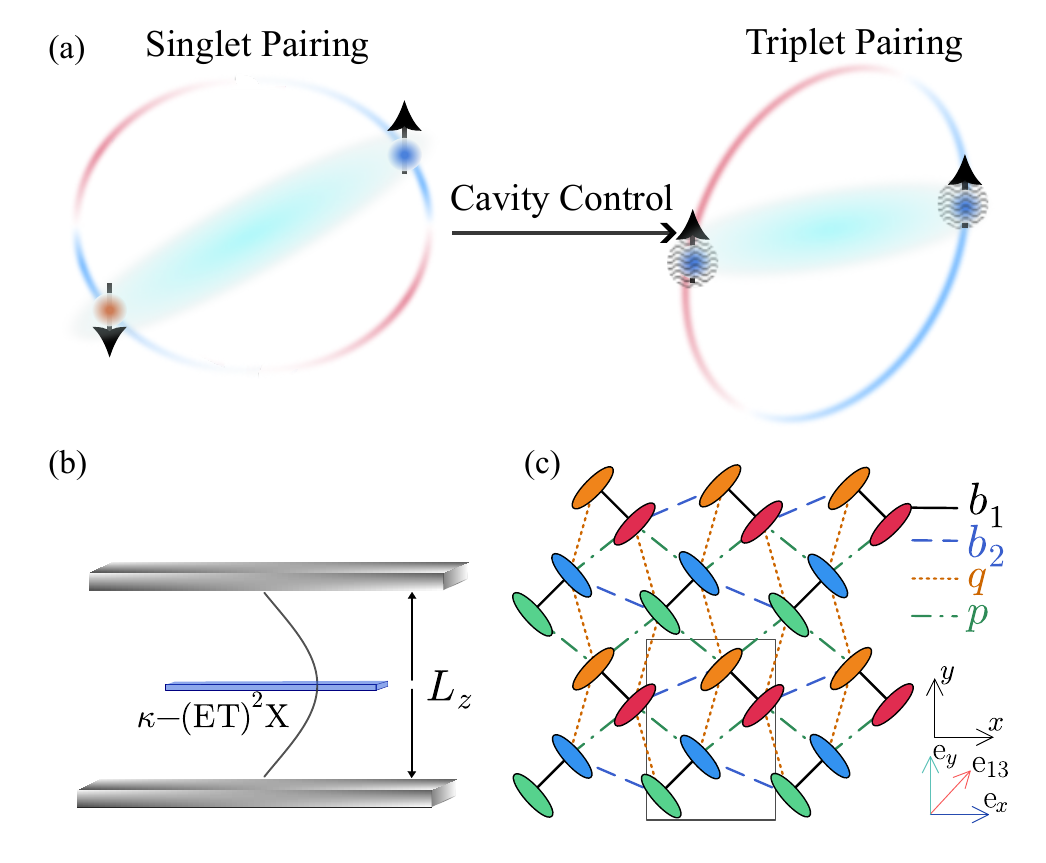}
    \caption{(a) Schematic of cavity-controlled switching from a $d_{xy}$-wave singlet to a $p$-wave triplet channel. (b) $\kappa$-(ET)$_2$X crystal embedded in a cavity with size $L_z$ along the $z$ direction. (c) Quasi-two-dimensional $\kappa$-(ET)$_2$X lattice with four bond types defining the bare hoppings $t_{ij}$. Blue, orange, green, and red ellipses label the four molecular sites $l=1$--$4$ in each unit cell, located at $\mathbf r_{1}/a=(-0.1,-0.2)$, $\mathbf r_{2}/a=(0.1,0.5)$, $\mathbf r_{3}/a=(-0.4,-0.5)$, and $\mathbf r_{4}/a=(0.4,0.2)$, where $a=0.8\,\mathrm{nm}$ is the lattice constant along the $x$ direction. Blue, green, and red arrows indicate the three cavity polarization directions considered here: $\mathbf e_x=(1,0)$, $\mathbf e_y=(0,1)$, and $\mathbf e_{13}=(1,1)/\sqrt{2}$.}
    \label{setup}
\end{figure}

Recent dark-cavity experiments have reported measurable superconducting responses, including shifts of the transition temperature and suppression of the superfluid density~\cite{Thomas2021,Thomas2025,Keren2025}. These findings place dark-cavity superconductivity in a regime where the cavity vacuum field couples to material degrees of freedom without external pumping~\cite{Lu2025,Rokaj2022,Li2022}. Microscopically, cavity coupling can dress phonons into phonon polaritons~\cite{Sentef2018,Hagenmuller2019}, mediate effective electron-electron interactions~\cite{Schlawin2019,Andolina2024,Li2022}, or reconstruct the low-energy electronic structure~\cite{Lu2024,Kozin2025}. These developments raise the central question of whether cavity control can steer pairing symmetry in unconventional superconductors.

\begin{table*}[t]
\centering
\begin{minipage}[t]{0.44\textwidth}
\centering
\begin{tabular}{|c|c|c|}
\hline
$\Grep_{\mathrm{sub}}$ & even basis functions & odd basis functions \\
\hline
$A_1$ & $\mathbb Q^{0}_{A_1}$, $\mathbb Q^{12}_{A_1}$, $\mathbb Q^{13}_{A_1}$, $\mathbb Q^{14}_{A_1}$, $\mathbb Q^{23}_{A_1}$, $\mathbb Q^{34}_{A_1}$
     & $\mathbb T^{12}_{A_1}$, $\mathbb T^{14}_{A_1}$, $\mathbb T^{23}_{A_1}$, $\mathbb T^{34}_{A_1}$ \\
$A_2$ & $\mathbb Q^{0}_{A_2}$, $\mathbb Q^{12}_{A_2}$, $\mathbb Q^{13}_{A_2}$, $\mathbb Q^{14}_{A_2}$, $\mathbb Q^{23}_{A_2}$, $\mathbb Q^{34}_{A_2}$
     & $\mathbb T^{12}_{A_2}$, $\mathbb T^{14}_{A_2}$, $\mathbb T^{23}_{A_2}$, $\mathbb T^{34}_{A_2}$ \\
$B_1$ & $\mathbb Q^{0}_{B_1}$, $\mathbb Q^{12}_{B_1}$, $\mathbb Q^{14}_{B_1}$, $\mathbb Q^{23}_{B_1}$, $\mathbb Q^{34}_{B_1}$
     & $\mathbb T^{12}_{B_1}$, $\mathbb T^{13}_{B_1}$, $\mathbb T^{14}_{B_1}$, $\mathbb T^{23}_{B_1}$, $\mathbb T^{34}_{B_1}$ \\
$B_2$ & $\mathbb Q^{0}_{B_2}$, $\mathbb Q^{12}_{B_2}$, $\mathbb Q^{14}_{B_2}$, $\mathbb Q^{23}_{B_2}$, $\mathbb Q^{34}_{B_2}$
     & $\mathbb T^{12}_{B_2}$, $\mathbb T^{13}_{B_2}$, $\mathbb T^{14}_{B_2}$, $\mathbb T^{23}_{B_2}$, $\mathbb T^{34}_{B_2}$ \\
\hline
\end{tabular}
\end{minipage}
\hfill
\begin{minipage}[t]{0.06\textwidth}
\centering
{~~~~~~~~~~~~~~~\Large$\otimes$}
\end{minipage}
\hfill
\begin{minipage}[t]{0.44\textwidth}
\centering
\begin{tabular}{|c|c|}
\hline
$\Grep_{\mathbf{k}}$ & basis functions \\
\hline
$A_{1}$ & $1$, $\cos(k_xa)$, $\cos(k_yb)$, $\cos(k_xa)\cos(k_yb)$ \\
$A_{2}$ & $\sin(k_xa)\sin(k_yb)$ \\
$B_{1}$ & $\sin(k_xa)$, $\sin(k_xa)\cos(k_yb)$ \\
$B_{2}$ & $\sin(k_yb)$, $\cos(k_xa)\sin(k_yb)$ \\
\hline
\end{tabular}
\end{minipage}
\caption{Basis functions used to construct the symmetry-adapted gap matrices.}
\label{table1}
\end{table*}

Among unconventional superconductors, stabilizing spin-triplet pairing remains a longstanding goal because it offers a route to odd-parity and topological superconductivity~\cite{Sigrist1991,Sato2017}. Established routes include Zeeman depairing~\cite{Clogston1962,Chandrasekhar1962}, ferromagnetic spin fluctuations~\cite{BerkSchrieffer1966,FayAppel1980}, Fermi-surface (FS) tuning via strain~\cite{Hicks2014}, and engineered superconductor-ferromagnet heterostructures~\cite{Bergeret2005,LinderRobinson2015}. 
Although cavity-mediated electron-electron interactions have been proposed as a route to Amperean triplet pairing in certain settings~\cite{Schlawin2019}, this scenario appears to depend on rather specific conditions and may be significantly weakened for non-parabolic dispersions~\cite{Andolina2024}.
This motivates a complementary question of whether a cavity can instead reshape the competition among existing pairing channels strongly enough to make a triplet state the leading instability in a realistic material~\cite{Mandal2023}, as sketched in Fig.~\ref{setup}(a).

Here we show that a dark cavity alters pairing tendencies through vacuum-induced band renormalization. In $\kappa$-(ET)$_2$X, cavity dressing anisotropically reshapes the electronic dispersion, enhances the low-energy density of states, and redistributes the phase space entering the linearized Bardeen-Cooper-Schrieffer (BCS) equation. As a result, the dominant singlet $d_{xy}$-wave instability is suppressed and, for the diagonal polarization $\mathbf e_{13}$, a triplet $p$-wave channel becomes the leading instability beyond a critical coupling. This change leaves clear signatures in the gap structure and in the low-energy quasiparticle spectrum.

{\it Symmetry classification of pairing.--}
We consider $\kappa$-(ET)$_2$X embedded in a cavity [Fig.~\ref{setup}(b)]. The conducting layer is described by a tight-binding model with four inequivalent ET molecules per unit cell~\cite{Kino1996,Watanabe2017,Seo2004,Wolter2007,Strack2005} [Fig.~\ref{setup}(c)], so the superconducting order parameter is intrinsically multicomponent. In the molecular basis, $\hat c_{\mathbf k \sigma}=(c_{\mathbf k1\sigma},c_{\mathbf k2\sigma},c_{\mathbf k3\sigma},c_{\mathbf k4\sigma})^T$, where $c_{\mathbf k l \sigma}$ annihilates an electron with crystal momentum $\mathbf k$ and spin $\sigma=\uparrow,\downarrow$ on molecule $l=1,\dots,4$ in the unit cell. The pairing matrix $\Delta_{ll^{\prime}}^{\sigma\sigma'}(\mathbf k)$, with fermionic antisymmetry
\begin{equation}
\Delta_{ll^{\prime}}^{\sigma\sigma'}(\mathbf{k})=-\Delta_{l^{\prime}l}^{\sigma'\sigma}(-\mathbf{k}).
\label{eq:fermionic_antisym}
\end{equation}
Because the cavity induces neither spin-orbit coupling nor a Zeeman field, the normal state preserves SU(2) spin symmetric. The gap matrix therefore decomposes into spin-singlet and spin-triplet channels. Suppressing the explicit spin structure, the corresponding reduced matrices $\Delta^{(s)}_{ll^{\prime}}(\mathbf k)$ and $\Delta^{(t)}_{ll^{\prime}}(\mathbf k)$ obey
\begin{equation}
\Delta^{(s)}_{ll^{\prime}}(\mathbf k)=\Delta^{(s)}_{l^{\prime}l}(-\mathbf k),\qquad
\Delta^{(t)}_{ll^{\prime}}(\mathbf k)=-\Delta^{(t)}_{l^{\prime}l}(-\mathbf k).
\label{eq:stpair}
\end{equation}

The crystal belongs to the nonsymmorphic layer group $p2gg$ ($\#8$), generated by translations, a twofold rotation $C_2$, and two glide reflections $g_x$ and $g_y$. In the four-sublattice Bloch basis, a symmetry operation $g$ acts through a momentum-dependent matrix $U_g(\mathbf k)$, so that $\Delta(\mathbf k)\to U_g(\mathbf k)\,\Delta(g^{-1}\mathbf k)\,U_g^{\mathsf T}(-\mathbf k)$.
Factoring out translations yields the point group $C_{2v}$, with irreducible representations (irreps) $\Grep=A_1,A_2,B_1,B_2$~\cite{Hayami2020}. Projecting trial pairing matrices onto these irreps gives the symmetry-adapted channels listed in Table~\ref{table1}~\cite{Platt2013} [see  Supplemental Materials (SM)~\cite{SM}].

Within a fixed spin sector, we expand
\begin{equation}
\Delta^{(s/t)}(\mathbf{k})
=
\sum_n \eta_n^{(s/t)}\,\Phi_n^{(s/t)}(\mathbf{k}),
\qquad
n\equiv(\Grep,\beta),
\label{eq:gap_channel_expand}
\end{equation}
where $\Phi_n^{(s/t)}(\mathbf{k})$ are matrix-valued basis functions in sublattice space. For compactness, Table~\ref{table1} organizes the basis functions as
\begin{equation}
\Phi_{\Grep,\beta}^{(s/t)}(\mathbf{k})
=
f_{\Grep_{\mathbf{k}},\gamma}(\mathbf{k})\,
\mathbb M_{\Grep_{\mathrm{sub}},\tau}(\mathbf{k}),
\qquad
\beta\equiv(\gamma,\tau),
\label{eq:factorized_basis}
\end{equation}
with scalar form factor $f_{\Grep_{\mathbf{k}},\gamma}(\mathbf{k})$ in the irrep $\Grep_{\mathbf{k}}$ and bond matrix $\mathbb M_{\Grep_{\mathrm{sub}},\tau}(\mathbf{k})$ in the irrep $\Grep_{\mathrm{sub}}$, where $\gamma$ and $\tau$ label the corresponding basis functions. $\mathbb M$ is taken from either $\mathbb Q$ or $\mathbb T$. The full irrep is
$\Grep=\Grep_{\mathbf{k}}\otimes\Grep_{\mathrm{sub}}$,
and the bond matrices satisfy
$\mathbb Q_{\Grep_{\mathrm{sub}},\tau}(\mathbf{k})
=
\mathbb Q_{\Grep_{\mathrm{sub}},\tau}^{\mathsf T}(-\mathbf{k})$
and
$\mathbb T_{\Grep_{\mathrm{sub}},\tau}(\mathbf{k})
=
-\,\mathbb T_{\Grep_{\mathrm{sub}},\tau}^{\mathsf T}(-\mathbf{k})$.
Combined with the parity of $f_{\Grep_{\mathbf{k}},\gamma}(\mathbf{k})$, this gives
$\Phi_{\Grep,\beta}^{\mathsf T}(\mathbf{k})=\pm\Phi_{\Grep,\beta}(-\mathbf{k})$,
with $+$ for singlet and $-$ for triplet channels. Here $a$ and $b$ are the lattice constants, and
$\phi_x=e^{i\mathbf{k}\cdot\mathbf{a}}$ and
$\phi_y=e^{i\mathbf{k}\cdot\mathbf{b}}$
are Bloch phases for bonds crossing the primitive vectors $\mathbf{a}$ and $\mathbf{b}$.

To make the matrix structure explicit, we choose representative basis functions with $f(\mathbf{k})=1$, so that the full basis function is determined entirely by bond matrix part. For example, the singlet basis function associated with $\mathbb Q_{A_2}^{14}$ is
\begin{equation}
\Phi^{(s)}_{A_2}(\mathbf{k})=
\begin{pmatrix}
0&0&0&1-\phi_x\\
0&0&\phi_y^*(\phi_x^*-1)&0\\
0&\phi_y(\phi_x-1)&0&0\\
1-\phi_x^*&0&0&0
\end{pmatrix},
\label{eq:Phi_A2_s}
\end{equation}
with nonzero entries only on the $(1,4)$ and $(2,3)$ bonds, corresponding to the $p$ links in Fig.~\ref{setup}(c). The constant terms represent intracell pairing, while the phase factors $\phi_x$, $\phi_y$, and their products encode the corresponding intercell components in neighboring unit cells. Likewise, the triplet basis function associated with $\mathbb T_{B_2}^{14}$ is
\begin{equation}
\Phi^{(t)}_{B_2}(\mathbf{k})=
\begin{pmatrix}
0&0&0&1+\phi_x\\
0&0&\phi_y^*(1+\phi_x^*)&0\\
0&-\phi_y(1+\phi_x)&0&0\\
-1-\phi_x^*&0&0&0
\end{pmatrix},
\label{eq:Phi_B2_t}
\end{equation}
with support on the same bonds but a different relative sign structure. Because both $\Phi^{(s)}_{A_2}$ and $\Phi^{(t)}_{B_2}$ are built from the $(1,4)$ and $(2,3)$ bonds associated with $t_p$, anisotropic cavity dressing of these links can influence both channels. 

The symmetry analysis is purely kinematic. It identifies the symmetry-allowed pairing channels and their associated real-space bond structures, but it does not determine which channel becomes the leading instability. Resolving this competition requires specifying the cavity-dressed Hamiltonian. Because singlet and triplet gap functions carry distinct bond and momentum-space structures, cavity-induced hopping renormalization and the resulting FS anisotropy reshape the pairing kernel in a channel-dependent manner and can thereby favor triplet pairing energetically.

{\it Band-structure renormalization.--}
To determine which pairing sector is favored, we next examine how cavity dressing renormalizes the normal-state dispersion. This band-structure renormalization modifies the FS and Fermi velocity $v_F$ entering the Cooper kernel and thus shifts the energetic balance between singlet and triplet channels.

We incorporate the cavity field through a gauge-invariant Peierls phase in the hopping, so that the quantized cavity mode dresses electron tunneling and renormalizes the bare hopping amplitudes~\cite{Li2022}. In the high-frequency regime, a quantum-Floquet expansion followed by projection onto the zero-photon sector yields 
\begin{equation}
H_0=-\sum_{ij,\sigma}\tilde t_{ij}\bigl(c^\dagger_{i\sigma}c_{j\sigma}+{\rm H.c.}\bigr),
\end{equation}
where $\tilde t_{ij}$ denotes the cavity-dressed hopping amplitude between sites $i$ and $j$. The corresponding Bloch Hamiltonian is $\hat h(\mathbf{k})$, with dispersions $\varepsilon_\nu(\mathbf{k})$ for $\nu=1,\dots,4$, where $\nu$ labels the four bands. We then define
$\xi_\nu(\mathbf{k})=\varepsilon_\nu(\mathbf{k})-\mu$,
where $\mu$ is fixed by the electron density $n_{\mathrm{el}}=3/2$ per molecular site, corresponding to $3/4$ filling in $\kappa$-(ET)$_2$X~\cite{Powell2006}. 
The cavity dressing is parametrized by an effective light-matter coupling $g_{\mathrm{eff}}$, which is treated in this work as a tunable parameter. Experimentally,
$g_{\mathrm{eff}}$ can be controlled through cavity properties such as the frequency $\omega_c$ and compression factor $\mathcal{C}$.
(Appendix~\ref{appendixA}).

\begin{figure}[htbp]
    \centering
    \includegraphics[width=0.5\textwidth]{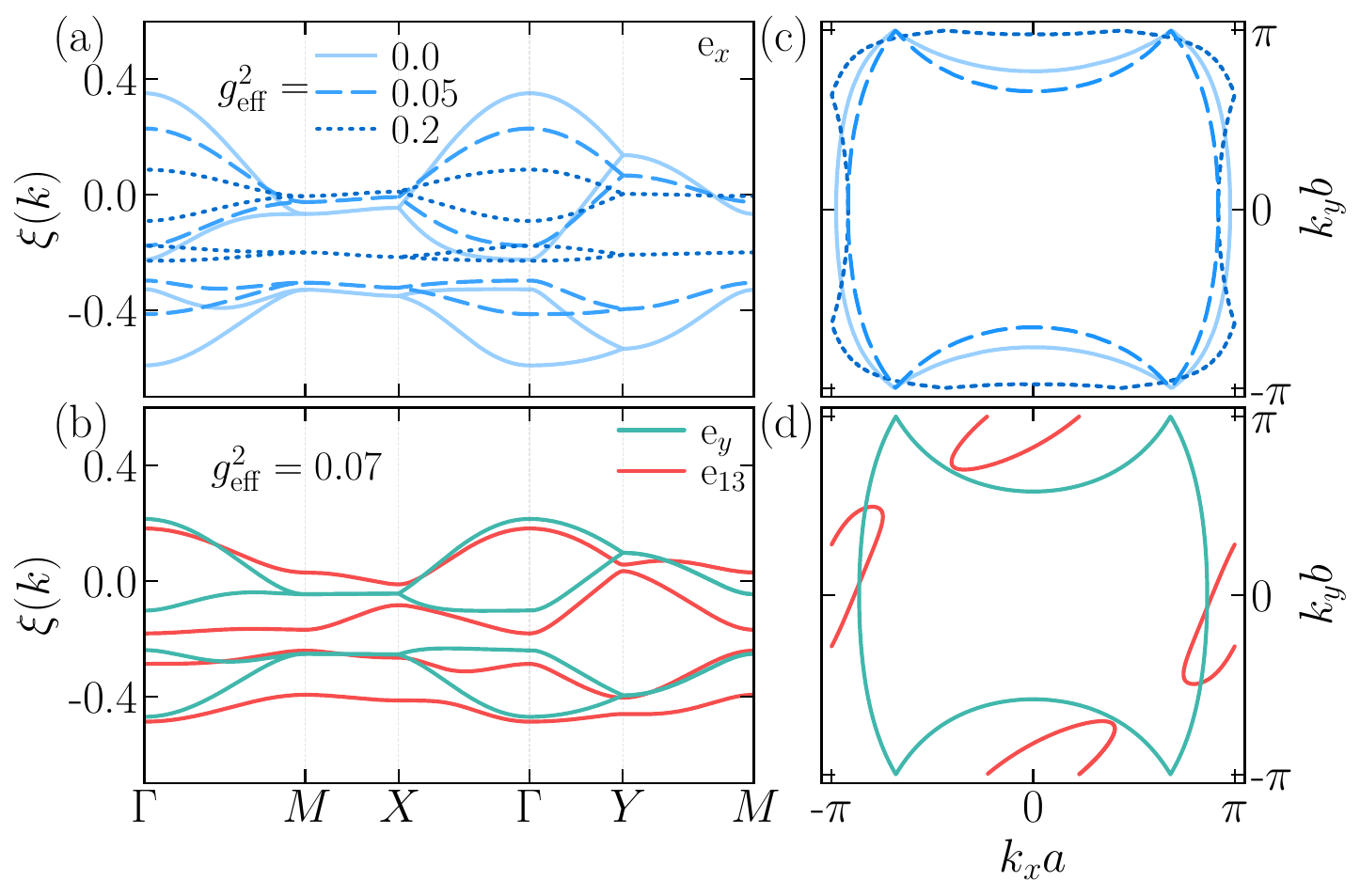}
    \caption{(a) Dispersion $\xi_\nu(\mathbf{k})$ at zero temperature along the high-symmetry path for a cavity polarized along $\mathbf e_x$, comparing the cavity-free case (solid) with $g_{\mathrm{eff}}^2=0.05$ (dashed) and $g_{\mathrm{eff}}^2=0.2$ (dotted). (b) Dispersion $\xi_\nu(\mathbf{k})$ for cavities polarized along $\mathbf e_y$ (green) and $\mathbf e_{13}=(1,1)/\sqrt{2}$ (red) at $g_{\mathrm{eff}}^2=0.07$. (c) and (d) Corresponding FS contours for panels (a) and (b), respectively.}
    \label{bands_cavity}
\end{figure}

In the cavity-free limit, the bands transform as one-dimensional irreps of $C_{2v}$. Nonsymmorphic symmetry enforces glide-protected band sticking along $M\!\to\!X$ and $M\!\to\!Y$, giving doublets that can be organized as $A_1\oplus A_2$ and $B_1\oplus B_2$. Cavity dressing narrows the bandwidth for all polarizations but does so anisotropically. For $\mathbf e_x$, the dispersion near the Fermi level is strongly flattened along $Y\!\to\!M$, with weaker changes along $X\!\to\!M$ [Fig.~\ref{bands_cavity}(a)]; for $\mathbf e_y$, the trend is reversed [green curves in Fig.~\ref{bands_cavity}(b)]. In both cases, the glide-protected degeneracies along $M\!\to\!X$ and $M\!\to\!Y$ remain intact.

The anisotropy follows directly from the polarization dependence of $\tilde t_{ij}$: bonds more nearly aligned with the cavity polarization $\mathbf e$ are more strongly suppressed. Thus $\mathbf e_x$ mainly reduces the intercell hopping $t_{b_2}$, enhancing the flattening along $Y\!\to\!M$, whereas $\mathbf e_y$ predominantly suppresses $t_q$, enhancing the flattening along $X\!\to\!M$. The corresponding FS sheets are shown in Fig.~\ref{bands_cavity}(c) and (d). Starting from an approximately elliptical contour in the cavity-free case, the FS becomes elongated mainly along $k_x$ for $\mathbf e_x$ and along $k_y$ for $\mathbf e_y$.

For $\mathbf e_{13}=(1,1)/\sqrt{2}$, the behavior changes qualitatively. This polarization breaks both glide symmetries $g_x$ and $g_y$, lifts the glide-protected degeneracies, and opens gaps along the corresponding high-symmetry lines. Microscopically, the two intracell $b_1$ bonds are dressed unequally because $\mathbf e_{13}$ is parallel to one bond but perpendicular to the other. The FS is then reconstructed into two disconnected contours [red curves in Fig.~\ref{bands_cavity}(d)]. These cavity-induced changes in the dispersion and FS directly renormalize the Fermi velocity anisotropy entering the pairing kernel.

{\it Linearized BCS theory.--}
Projecting the gap equation onto the symmetry-adapted basis in Eq.~(\ref{eq:gap_channel_expand}) yields the linearized eigenvalue problem near $T_c$~\cite{Sigrist1991,Scalapino2012} (Appendix \ref{appendixB})
\begin{equation}
\lambda\,\eta_n^{(s/t)}
=
\sum_{n'} \mathbb K_{nn'}^{(s/t)}\,\eta_{n'}^{(s/t)},
\qquad
\mathbb K^{(s/t)}=\mathbb V^{(s/t)}\mathbb C^{(s/t)},
\label{eq:kernel}
\end{equation}
where $\mathbb V^{(s/t)}$ is the interaction projected onto the symmetry basis and $\mathbb C^{(s/t)}$ is the corresponding Cooper matrix. In the linearized regime, $\mathbb C^{(s/t)}$ is dominated by low-energy states near the FS and weighted by $1/v_F$.

Equivalently, in the band representation and neglecting interband Cooper pairing, the instability can be written in FS-projected form as
\begin{eqnarray}
\lambda\,\Delta_\nu^{(s/t)}(\mathbf{k})
&=&
-\sum_{\nu'}
\int 
\oint_{\mathrm{FS}_{\nu'}}
\frac{d\xi'dS_{\mathbf{k}'}}{(2\pi)^2\,v_{F,\nu'}(\mathbf{k}')}V_{\nu\nu'}^{(s/t)}(\mathbf{k},\mathbf{k}')
\nonumber\\
&&\qquad\qquad\times\frac{\tanh(\xi'/(2T))}{2\xi'}\,
\Delta_{\nu'}^{(s/t)}(\mathbf{k}'),
\label{eq:linearized_gap}    
\end{eqnarray}
where $\Delta_\nu^{(s/t)}(\mathbf{k})$ is the gap on FS sheet $\nu$, $V_{\nu\nu'}^{(s/t)}$ is the interaction projected to the FS, and $v_{F,\nu}$ is the Fermi velocity. Here the intraband approximation means that each Cooper pair is formed within a single band, while pair scattering between different FS sheets is still retained through $V_{\nu\nu'}^{(s/t)}$. Equation~(\ref{eq:kernel}) follows by projecting Eq.~(\ref{eq:linearized_gap}) onto the basis $\Phi_n^{(s/t)}(\mathbf{k})$.

The leading pairing symmetry is set by the eigenvector with the largest eigenvalue $\lambda_{\max}$, and the transition temperature satisfies
$\lambda_{\max}(T_c)=1$.
Equation~(\ref{eq:linearized_gap}) shows that the pairing strength depends not only on the interaction vertex but also on the FS weight $1/v_F(\mathbf{k})$~\cite{Scalapino2012}. A uniform reduction of $v_F$ would enhance all channels similarly, whereas cavity-induced bond-selective renormalization enhances $1/v_F$ only on specific FS patches. The relative stability of competing pairing states is therefore controlled by how strongly their symmetry form factors are concentrated on those patches (see SM~\cite{SM}).

{\it Cavity-reweighted pairing instabilities.--}
We now ask whether cavity vacuum dressing can actively promote triplet superconductivity. To this end, we consider the interaction part of the effective extended Hubbard model~\cite{Seo2000,Kino1996},
\begin{equation}
H_{\rm int}
=
U\sum_i n_{i\uparrow}n_{i\downarrow}
+\sum_{\langle i,j\rangle}V_{ij}\,n_i n_j,
\label{eq:Hint}
\end{equation}
where $n_{i\sigma}=c_{i\sigma}^\dagger c_{i\sigma}$ and $n_i=n_{i\uparrow}+n_{i\downarrow}$. Projecting Eq.~(\ref{eq:Hint}) onto the Cooper channel formed by states at $\mathbf k$ and $-\mathbf k$ in the truncated basis of Table~\ref{table1} yields the interaction matrix $\mathbb V$ entering Eq.~(\ref{eq:kernel}). In this basis, the on-site term $U$ contributes only to the even-parity spin-singlet on-site form factor, whereas the nearest-neighbor interaction $V_{ij}$ carries nontrivial momentum dependence and contributes to both singlet and triplet channels (see SM~\cite{SM}).

\begin{figure}[htbp]
    \centering
    \includegraphics[width=0.5\textwidth]{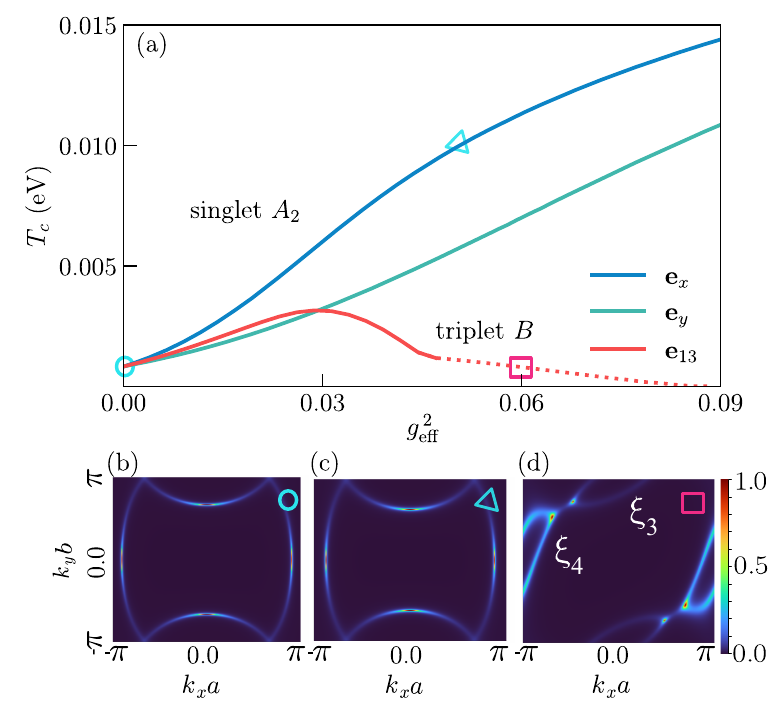}
    \caption{(a) Superconducting transition temperature $T_c$ from linearized BCS theory as a function of $g_{\mathrm{eff}}^2$ for cavities polarized along $\mathbf e_x$ (blue), $\mathbf e_y$ (green), and $\mathbf e_{13}$ (red). Along each curve, solid and dotted segments indicate parameter regimes where the dominant pairing channel is singlet or triplet, respectively, and the corresponding irrep is indicated in the panel. The points marked in (a) correspond to panels (b)--(d), which display the normalized zero-energy spectral function $\tilde{\mathcal A}(\mathbf{k},\omega=0)$ for the indicated cases.}
    \label{phase}
\end{figure}

Figure~\ref{phase}(a) shows that the decisive role of the cavity is not merely to enhance the overall pairing scale, but to reweight the competing pairing channels differently along the FS. A weak cavity generally increases $T_c$ by narrowing the bandwidth and reducing the Fermi velocity, thereby enhancing the low-energy weight entering $\mathbb C$~\cite{Kozin2025}. However, the central effect is the polarization-dependent redistribution of this weight. Since singlet and triplet form factors draw support from different FS regions, the cavity can selectively weaken one channel relative to the other.

The contrast between a uniform enhancement of pairing and a channel-selective reweighting of the pairing kernel is most pronounced for $\mathbf e_{13}$ polarization. In that case, the cavity breaks the glide symmetries and lowers the spatial symmetry, reducing the pairing classification to two sectors. In the language of $C_{2v}$, the lowered symmetry allows mixing within $(A_1,A_2)$ and within $(B_1,B_2)$. At the same time, $\mathbf e_{13}$ produces a strong momentum-dependent modulation of $v_F(\mathbf k)$ that suppresses the FS weight on the segments contributing most strongly to the dominant singlet channel. Together, symmetry lowering and momentum-selective reweighting reshuffle the pairing eigenvalues and, beyond a critical coupling, drive the leading instability from singlet to triplet pairing. By contrast, for $\mathbf e_x$ and $\mathbf e_y$ the FS regions favoring the singlet state remain largely intact, so the leading instability stays in the singlet $A_2$ channel throughout the coupling range shown. A quantitative analysis of this redistribution is given in the SM~\cite{SM}.

The resulting gap evolution is visualized through the normalized zero-energy spectral function $\tilde{\mathcal A}(\mathbf k,\omega=0)$ in Figs.~\ref{phase}(b)-\ref{phase}(d). In the cavity-free case, Fig.~\ref{phase}(b) shows the characteristic nodal structure of a singlet $d_{xy}$-wave superconductor, consistent with earlier reports~\cite{Watanabe2019,Guterding2016,Mizukami2025,Malone2010}. The $\mathbf e_x$ polarization, and similarly $\mathbf e_y$, preserves this overall nodal pattern, consistent with the persistence of the singlet $A_2$ instability. The $\mathbf e_{13}$ polarization is qualitatively different: as shown in Fig.~\ref{phase}(d), the FS splits into two disconnected contours associated with two bands, and each contour exhibits a two-node (time-reversal pair) structure, amounting to four nodes across the entire quasiparticle spectrum.
The two-node structure reflects the sign-flip of the pairing function [Eq.~\eqref{eq:stpair}] in the $p$-wave triplet superconducting state (see SM~\cite{SM}).

{\it Conclusion.--}
In summary, dark-cavity vacuum dressing offers a route to control superconductivity. Its central effect is to reshape the low-energy electronic structure near the FS. In particular, it changes the Fermi velocity and redistributes the weight of FS states in a polarization-dependent way. Within our linearized BCS analysis, this redistribution can enhance $T_c$ and can also drive the leading instability from singlet to triplet pairing. In the multicomponent system studied here, the same dressing reshapes the low-energy dispersion and leaves characteristic signatures in $\tilde{\mathcal A}(\mathbf k,\omega=0)$ that should be observable in momentum-resolved photoemission. Complementary bulk evidence could also come from low-temperature thermal-conductivity measurements~\cite{Matsuda2006}, which are sensitive to low-energy quasiparticles and can distinguish between nodal and nodeless realizations of the proposed triplet pairing state through their characteristic power-law and exponential temperature dependences, respectively.
Beyond these probes, THz near-field spectroscopy may provide additional access to low-energy collective excitations of the cavity-modified superconducting state~\cite{Sun2020}.

In $\kappa$-(ET)$_2$X, recent experiments have already observed cavity-induced changes in the superfluid density~\cite{Keren2025}, while earlier optical-pump studies have also reported light-induced changes in the superconducting response~\cite{Buzzi2020}. However, no change in pairing symmetry has been identified so far. Within our framework, this is naturally understood from the fact that current experiments still probe a relatively weak-coupling regime, $g_{\mathrm{eff}}\ll0.01$. In this regime, cavity dressing can modify superconducting properties, but it is not yet strong enough for the triplet channel to become dominant over the singlet one. We also emphasize that the mechanism proposed in Ref.~\cite{Keren2025} is conceptually distinct from the one considered here. In that work, hBN phonon modes weaken the phonon-mediated pairing glue in $\kappa-\mathrm{(ET)}_2\mathrm{X}$. In our setup, by contrast, vacuum electromagnetic fluctuations renormalize the underlying electronic dispersion and reshape the FS in a polarization-dependent way. This, in turn, changes the balance among the symmetry-allowed pairing channels. Our estimates indicate that reaching the triplet-dominant regime requires $a\omega_c/\pi\mathcal{C}\sim100$, corresponding to a cavity-field compression factor of $\mathcal{C}\sim10^{-6}$ for $\omega_c\sim 0.3\mathrm{eV}$. Although demanding, such confinement appears plausible in deep-subwavelength nanocavities~\cite{Sheinfux2024}. Additional tuning, for example by pressure or Floquet dressing, may also help place the system in the controlled regime of the quantum-Floquet expansion~\cite{Erkenov2024,Kandpal2009,Yonemitsu2017}. More broadly, cavity-induced symmetry reduction should be even more effective in higher-symmetry materials and moir'e superlattices~\cite{Lin2026}, where stronger light-matter effects may further amplify the reweighting of competing pairing channels.

{\it Acknowledgements.---}
This work is supported by the National Key R\&D Program of China (No. 2021YFA1400900, No. 2022YFA1404701, No. 2022YFA1404500, 24LZ1400900, No. 2024ZD0300100) and Shanghai Science and Technology Innovation Action Plan (No. 24Z510205936, 24DP2600100).

\newpage 

\begin{widetext} 
    \centering{\bf End Matter} 
\end{widetext}
\appendix
\setcounter{secnumdepth}{1}

\section{Details of the Model}\label{appendixA}
\setcounter{equation}{0}
\renewcommand{\theequation}{A.\arabic{equation}}

In the absence of the cavity, electron motion is described by nearest-neighbor hoppings $t_{ij}$ on the four distinct bonds $(b_1,b_2,p,q)$ of the $\kappa$-type geometry~\cite{Kino1996,Watanabe2017}, as shown in Fig.~\ref{setup}(c). In the cavity, the kinetic sector couples to the quantized electromagnetic field through the gauge-invariant Peierls substitution, so that
\begin{equation}
H = H_0 + U\sum_i \hat n_{i\uparrow}\hat n_{i\downarrow}
+\sum_{\langle i,j\rangle}V_{ij}\,\hat n_i \hat n_j,
\label{H_total}
\end{equation}
where $i,j$ label molecular sites in real space, $\langle i,j\rangle$ denotes nearest-neighbor bonds, $\hat n_{i\sigma}=\hat c_{i\sigma}^\dagger \hat c_{i\sigma}$, and $\hat n_i=\hat n_{i\uparrow}+\hat n_{i\downarrow}$. The cavity-coupled quadratic part is
\begin{equation}
H_0=
-\sum_{\langle ij\rangle,\sigma}
\left(
t_{ij}\,e^{i\hat\chi_{ij}}\hat c^\dagger_{i\sigma}\hat c_{j\sigma}
+{\rm H.c.}
\right)
+\sum_{\mathbf q}\omega_{\mathbf q}\hat a^\dagger_{\mathbf q}\hat a_{\mathbf q},
\label{H0}
\end{equation}
where $\hat a_{\mathbf q}$ and $\hat a^\dagger_{\mathbf q}$ annihilate and create a cavity photon of frequency $\omega_{\mathbf q}$, and $\hat\chi_{ij}\approx e\,\mathbf A(\mathbf R_{ij})\cdot\mathbf d_{ij}$, with bond center $\mathbf R_{ij}=(\mathbf r_i+\mathbf r_j)/2$ and bond vector $\mathbf d_{ij}=\mathbf r_i-\mathbf r_j$. For the planar cavity considered here, the in-plane photon momentum is continuous, while the out-of-plane component is quantized as $q_c=\pi/L_z$. We use the bare hopping parameters of deuterated $\kappa$-(ET)$_2$Cu[N(CN)$_2$]Br,
\begin{equation}
(t_{b_1},t_{b_2},t_p,t_q)=(196,65,105,-39)\,\mathrm{meV},
\end{equation}
taken from first-principles band calculations~\cite{Koretsune2014}.

Expanding $H_0$ in the photon Fock basis $|\mathbf n\rangle\equiv|\{n_{\mathbf q}\}\rangle$, with $\mathbf n=\{n_{\mathbf q}\}$ the photon occupation vector, we define the quantum-Floquet matrix elements
\begin{eqnarray}
\mathcal H_{\mathbf n,\mathbf m}
&=&\langle \mathbf n|H_0|\mathbf m\rangle
-\delta_{\mathbf n\mathbf m}\sum_{\mathbf q}n_{\mathbf q}\omega_{\mathbf q}\nonumber\\
&=&-\sum_{\langle ij\rangle,\sigma}
t_{ij}\,\langle \mathbf n|e^{i\hat\chi_{ij}}|\mathbf m\rangle\,
\hat c^\dagger_{i\sigma}\hat c_{j\sigma}
+{\rm H.c.},
\label{eq:qfloquet}
\end{eqnarray}
where $\mathbf m=\{m_{\mathbf q}\}$ is another photon occupation vector, $\delta_{\mathbf n\mathbf m}$ is the Kronecker delta, and $\mathbf 0$ denotes the photon vacuum. The Peierls matrix element factorizes over photon modes (see SM~\cite{SM}). In the high-frequency regime, projecting onto the vacuum sector $\mathbf 0$ gives
\begin{equation}
H_{\mathrm{eff}}^{(0)}=\mathcal H_{\mathbf 0,\mathbf 0}
=
-\sum_{ij,\sigma}t_{ij}^{(0)}
\bigl(\hat c^\dagger_{i\sigma}\hat c_{j\sigma}+{\rm H.c.}\bigr),
\end{equation}
with renormalized hopping
\begin{equation}
t_{ij}^{(0)}
=
t_{ij}e^{-2\Lambda |z_{ij}|^2 g_{\mathrm{eff}}^2},
\end{equation}
where
\begin{equation}
\Lambda=\sqrt{1+\left(\frac{Q_\parallel}{q_c}\right)^2}-1,
\qquad
z_{ij}=\frac{\mathbf e\cdot\mathbf d_{ij}}{a},
\end{equation}
with $Q_\parallel$ the in-plane momentum cutoff, and $\mathbf e$ the cavity polarization vector. The effective light-matter coupling is
\begin{equation}
g_{\mathrm{eff}}=\frac{a}{2L_z}\sqrt{\frac{\pi\alpha}{\mathcal C}},
\end{equation}
where $\alpha$ is the fine-structure constant and $\mathcal C$ is the cavity-field compression factor.

Virtual excursions into excited photon sectors generate higher-order corrections,
\begin{equation}
H_{\mathrm{eff}}^{(1)}
=
-\sum_{\mathbf n\neq\mathbf 0}
\frac{\mathcal H_{\mathbf 0,\mathbf n}\mathcal H_{\mathbf n,\mathbf 0}}
{\mathbf n\cdot\boldsymbol\omega},
\end{equation}
which we retain only at the level of induced quadratic hopping terms,
\begin{equation}
H_{\mathrm{eff}}^{(1)}
\approx
-\sum_{ij,\sigma}t_{ij}^{(1)}
\bigl(\hat c^\dagger_{i\sigma}\hat c_{j\sigma}+{\rm H.c.}\bigr).
\end{equation}
As shown in the SM~\cite{SM}, the dominant dressed hoppings $t_{ij}^{(0)}$ exceed the higher-order corrections $t_{ij}^{(1)}$ by more than two orders of magnitude, thereby justifying the effective Hamiltonian used in the main text.

We take the interaction sector to be the extended Hubbard term
\begin{equation}
H_{\rm int}
=
U\sum_i n_{i\uparrow}n_{i\downarrow}
+\sum_{\langle i,j\rangle}V_{ij}\,n_i n_j ,
\end{equation}
where $V_{ij}$ takes the values $(V_{b_1},V_{b_2},V_p,V_q)$ on the four bonds shown in Fig.~\ref{setup}(c). In the calculations below we use the representative parameter set
\begin{eqnarray}
U&=&0.23t_{b_1},\nonumber\\
(V_{b_1},V_{b_2},V_p,V_q)&=&(1,0.86,-0.27,-0.1)U.
\end{eqnarray}
These are effective couplings that already include renormalization from spin fluctuations rather than bare microscopic values. Cavity-induced corrections to the interaction vertex are expected to be of the same order as the neglected higher-order hopping terms and are therefore omitted throughout~(see SM~\cite{SM}).

\section{Linearized Gap Equation}\label{appendixB}
\setcounter{equation}{0}
\renewcommand{\theequation}{B.\arabic{equation}}

Within a fixed singlet or triplet sector, we suppress the superscript $(s/t)$ throughout and define the channel pair operators
\begin{eqnarray}
\hat\Delta_n^\dagger
&=&\frac{1}{\sqrt{N_k}}\sum_{\mathbf k}
\hat{\mathbf c}_{\mathbf k}^\dagger\,\Phi_n(\mathbf k)\,
\hat{\mathbf c}_{-\mathbf k}^{\dagger\,\mathsf T},
\nonumber\\
\hat\Delta_n
&=&\frac{1}{\sqrt{N_k}}\sum_{\mathbf k}
\hat{\mathbf c}_{-\mathbf k}^{\mathsf T}\,
\Phi_n^\dagger(\mathbf k)\,\hat{\mathbf c}_{\mathbf k},
\end{eqnarray}
where $N_k$ is the number of sampled momenta in the Brillouin zone. In this basis, the pairing part of the interaction takes the form
\begin{equation}
H_{\rm int}^{(\rm pair)}
=
\frac{1}{2}\sum_{nm}\mathbb V_{nm}\,\hat\Delta_n^\dagger\hat\Delta_m,
\end{equation}
where $\mathbb V_{nm}$ is the interaction matrix obtained by projecting the microscopic pairing vertex onto the symmetry-adapted basis $\{\Phi_n\}$.

Within the Bogoliubov--de Gennes mean-field approximation, the gap matrix is expanded as in Eq.~(\ref{eq:gap_channel_expand}), with coefficients
\begin{equation}
\eta_n=\sum_m\mathbb V_{nm}\,\langle\hat\Delta_m\rangle .
\label{eq:eta-self}
\end{equation}
Away from $T_c$, Eq.~(\ref{eq:eta-self}) is nonlinear because $\langle\hat\Delta_m\rangle$ depends on the full set $\{\eta_n\}$.

Near $T_c$, the order parameter is small, and
\begin{equation}
\langle\hat\Delta_m\rangle
=
\sum_\ell \mathbb C_{m\ell}(T)\,\eta_\ell,
\end{equation}
where $\mathbb C(T)$ is the normal-state Cooper matrix (see SM~\cite{SM}). Substituting this relation into Eq.~(\ref{eq:eta-self}) yields
\begin{equation}
\eta_n
=
\sum_{m\ell}\mathbb V_{nm}\,\mathbb C_{m\ell}(T)\,\eta_\ell
=
\sum_\ell \mathbb K_{n\ell}(T)\,\eta_\ell,
\label{eq:lin-gap}
\end{equation}
with
\begin{equation}
\mathbb K(T)=\mathbb V\,\mathbb C(T).
\end{equation}
The leading superconducting instability is determined by the largest eigenvalue $\lambda_{\max}(T)$ of $\mathbb K(T)$, and $T_c$ is fixed by $\lambda_{\max}(T_c)=1$.
\end{document}